\documentclass[
    ,final            
  ,numberedheadings 
  ]
  {aipproc}

\usepackage{theorem,amsmath,amssymb,mathrsfs}
\usepackage{float,supertabular}
\usepackage{eucal}

\layoutstyle{8x11single}

\def\index#1{}
\def\<{\langle}\def\>{\rangle}
\def\:{\hbox{\bf :}}

\def\Reals{\mathbb R}\def\Cmplx{\mathbb C}
\def\map#1{{\mathscr{#1}}}
\def\vec#1{{\boldsymbol{#1}}}
\def\set#1{{\sf #1}}\def\alg#1{{\mathcal #1}}\def\aA{\alg{A}}
\def\aI{\alg{I}}

\def\transp#1{{#1}^\tau}

\def\Op#1{\operatorname{Op}_{#1}}

\def\dim{\operatorname{dim}}\def\adm{\operatorname{adm}}
\def\idim#1{\operatorname{idim}(#1)}

\def\sH{\set{H}}

\def\qed{$\,\blacksquare$\par}
\def\eg{e. g. }\def\ie{i. e. }
\def\n#1{|\!|#1|\!|}\def\nn#1{|\!|\!|#1|\!|\!|}

\newtheorem{definition}{Definition}
\newtheorem{lemma}{Lemma}

\newtheorem{corollary}{Corollary}
\newtheorem{theorem}{Theorem}
\newtheorem{gaxiom}{General Axiom}
\newtheorem{postulate}{Postulate}
\newtheorem{grule}{Rule}
\def\Proof{\medskip\par\noindent{\bf Proof. }}
\theoremstyle{remark}\newtheorem{remark}{{\bf Remark}}
\theoremstyle{example}

\def\trnsfrm#1{\mathscr #1}
\def\tA{\trnsfrm A}\def\tB{\trnsfrm B}\def\tC{\trnsfrm C}\def\tS{\trnsfrm S}
\def\tI{\trnsfrm I}\def\tT{\trnsfrm T}\def\tM{\trnsfrm M}\def\tX{\trnsfrm X}

\def\AA{\mathbb A}\def\AB{\mathbb B}\def\AC{\mathbb C}\def\AL{\mathbb L}
\def\cA{{\underline{\tA}}}\def\cB{\underline{\tB}}\def\cI{{\underline{\tI}}}
\def\cX{{\underline{\tX}}}

\def\Stset{{\mathfrak S}}\def\Wset{{\mathfrak W}}
\def\Trnset{{\mathfrak T}}\def\trnset{{\mathfrak t}}
\def\Cntset{{\mathfrak P}}\def\Prdset{{\mathfrak P}_p}
\def\Idx#1{%
}

\begin{document}
\title{How to Derive the Hilbert-Space Formulation of Quantum Mechanics From Purely Operational
  Axioms\footnote{Work presented at the conference {\em On the Present Status of Quantum Mechanics}
    held on 7-9 September 2005, Mali Losinj, Croatia, in celebration of the 70th birthday of Gian
    Carlo Ghirardi. The dimensionality theorems along with all results already published in Ref.
    \cite{darianoVax2005} have been presented at the conference, whereas the operational definition
    of the real adjoint and the Hilbert spaces derivations have been presented as work in progress,
    and were completed two months later.}}  
\classification{03.65.-w} \keywords {Foundations,  Axiomatics, Measurement Theory} 
\author{Giacomo Mauro D'Ariano}{ address={{\em QUIT} Group,
    Dipartimento di Fisica ``A. Volta'', via Bassi 6,
    I-27100 Pavia, Italy, {\em http://www.qubit.it}\\
    Department of Electrical and Computer Engineering, Northwestern University, Evanston, IL 60208}}
\begin{abstract} In the present paper I show how it is possible to derive the Hilbert space
  formulation of Quantum Mechanics from a comprehensive definition of {\em physical experiment} and
  assuming {\em experimental accessibility and simplicity} as specified by five simple Postulates.
  This accomplishes the program presented in form of conjectures in the previous paper
  \cite{darianoVax2005}. Pivotal roles are played by the {\em local observability principle}, which
  reconciles the holism of nonlocality with the reductionism of local observation, and by the
  postulated existence of {\em informationally complete observables} and of a {\em symmetric
    faithful state}. This last notion allows one to introduce an operational definition for the real
  version of the ``adjoint"---i. e. the transposition---from which one can derive a real
  Hilbert-space structure via either the Mackey-Kakutani or the Gelfand-Naimark-Segal constructions.
  Here I analyze in detail only the Gelfand-Naimark-Segal construction, which leads to a real
  Hilbert space structure analogous to that of (classes of generally unbounded) selfadjoint
  operators in Quantum Mechanics. For finite dimensions, general dimensionality theorems that can be
  derived from a local observability principle, allow us to represent the elements of the real
  Hilbert space as operators over an underlying complex Hilbert space (see, however, a still open
  problem at the end of the paper). The route for the present operational axiomatization was
  suggested by novel ideas originated from Quantum Tomography.
\end{abstract}
\maketitle
\section{Introduction}
Quantum Mechanics is a sort of ``syntactic manual'' for physical theories: it is a set of rules that
hold for any physical field---electroweak, nuclear, gravitational---and apply to the entire physical
domain, from micro to macro, independently of the size and energy scale. Should we consider Quantum
Mechanics a General Law of Nature, or, instead, a Logical Necessity, a {\em Miniature Epistemology}?
Indeed, for the first time in the history of Physics, Quantum Mechanics in its very essence
addresses the crucial problem of the {\em Physical Measurement}, problem which is at the core of
Physics as an experimental science. It is not the physical description of the specific
instrumentation that I'm talking about, but the general process of information retrieval in {\em
  any} measurement, via interaction of the measured system with the measuring apparatus. I would say
that Quantum Mechanics more generally deals with the description of the {\em Physical Experiment},
which is indeed the epistemic archetype, the prototype {\em cognitive act} of interaction with
reality.

In the above framework it is mandatory to derive Quantum Mechanics from purely operational axioms.
This is not just for the sake of establishing more general and irreducible foundations, but also to
understand the intimate relations between general epistemic issues---such as locality, causality,
probability interpretations, holism versus reductionism, and growth of experimental complexity with
the ``size'' of the measured system.

In the present work the starting point for axiomatization is a very comprehensive definition of {\em
  physical experiment}. As I have shown in Ref. \cite{darianoVax2005}, the adoption of such a
general definition of experiment constitutes a very seminal point for axiomatization, entailing a
thorough series of notions that are usually considered of quantum nature---such as the same
probabilistic notion of state, and the notions of conditional state, local state, pure state,
faithful state, instrument, propensity (\ie "effect"), dynamical and informational equivalence,
dynamical and informational compatibility, predictability, discriminability, programmability,
locality, a-causality, and even many notions of dimensionality, orthogonality of states, rank of a
state, etc: for more details the interested reader is addressed to Ref.  \cite{darianoVax2005}. Here
we will see how, assuming {\em experimental accessibility and simplicity} in terms of five simple
operational axioms, the present conception of experiment brings his own Hilbert-space formulation,
which in turns entails the Quantum Mechanical one. The possibility of deriving the Hilbert-space
formulation from experimental simplicity/accessibility was first conjectured in the earlier attempt
\cite{darianoVax2005}.  As we will see, very interesting roles are played by Postulates numbered as
\ref{p:infocom}, \ref{p:locobs}, and \ref{p:faith} in the following, namely: (\ref{p:infocom}) the
assumed existence of informationally complete measurements, (\ref{p:locobs}) the local observability
principle, and (\ref{p:faith}) the existence of symmetric faithful states.  Postulate
\ref{p:infocom} minimizes the number of different apparatuses that are needed to retrieve any
different kind of information.  Postulate \ref{p:locobs} makes it possible to make joint
observations using only the same local measuring apparatuses used for measurements on single
systems. This also reconciles the holism of nonlocality with the reductionism of local observation.
Postulate \ref{p:faith} (in conjunction with the other two) allows one to calibrate any experimental
apparatus by just a single input state preparation.  It also allows one to introduce an operational
definition for the "real adjoint"---i. e. the transposition---from which one can derive a real
Hilbert space structure via either the Mackey-Kakutani \cite{Kakutani-Mackey} (see also Ref.
\cite{Istratescu}), or the Gelfand-Naimark-Segal \cite{GelfandNeumark} constructions.  Moreover, the
Postulates entail general dimensionality theorems, which are in agreement with the quantum
mechanical rule of tensor product of Hilbert spaces for composition of independent systems, and show
that the derived real Hilbert space is isomorphic to the real Hilbert space of Hermitian complex
matrices representing selfadjoint operators over a complex Hilbert space, which is the Hilbert space
formulation of Quantum Mechanics. In deriving one of the dimensionality theorems I have made,
however, the implicit assumption that the relation between the affine dimension and the
informational dimension for a convex set of state is the same for all physical systems---a sort of
informational universality (see the discussion at the end of the paper).

The present research has been stimulated by the recent noteworthy works on axiomatization of Quantum
Mechanics by L.  Hardy \cite{Hardyaxioms, HardyVax} and by C. Fuchs \cite{fuchs-vaxjo01}. However,
apart from a prominent role played by the informationally complete measurements, the relative
implications and connections with these works remain rather obscure to me, and will be object of
future studies.  Some expert readers will recognize strong affinities of the present work with the
program of G.  Ludwig \cite{Ludwig-axI}, who sought operational principles to select the structure
of quantum states from all possible convex structures (see also papers collected in the book
\cite{hartkamper74}). These works didn't have a followup mostly because the convex structure by
itself is quite poor mathematically. Here we use new crucial concepts that were almost unknown in
those years, concepts originated from the field of Quantum Tomography \cite{tomo_lecture}.  In
particular, recently it has been shown that it is possible to make a complete quantum calibration of
a measuring apparatus \cite{calib} or of a quantum operation \cite{tomo_channel} by using a single
pure bipartite state, and, more generally, using a faithful state \cite{faithful}. This gives us a
unique opportunity for deriving the Hilbert space structure from the convex structure in terms of
calibrability axioms, relying on the special link between the convex set of transformations and the
convex set of states which occurs in Quantum Mechanics, and which make the transformations of a
single system closely resemble the states of a bipartite system \cite{choi75,Jamiolkowski72}.

\section{The operational axiomatization}
\begin{gaxiom}[On experimental science]\label{ga:1}
  In any experimental science we make {\em experiments} to get {\em information} on the {\em state}
  of a {\em objectified physical system}. Knowledge of such a state will allow us to predict the
  results of forthcoming experiments on the same object system. Since we necessarily work with only
  partial {\em a priori} knowledge of both system and experimental apparatus, the rules for the
  experiment must be given in a probabilistic setting.
\end{gaxiom}
\begin{gaxiom}[On what is an experiment]\label{ga:2} An experiment on an
  object system consists in having it interact with an apparatus. The interaction between object
  and apparatus produces one of a set of possible transformations of the object, each one occurring
  with some probability. Information on the ``state'' of the object system at the beginning of the
  experiment is gained from the knowledge of which transformation occurred, which is the "outcome"
  of the experiment signaled by the apparatus.
\end{gaxiom}
\begin{postulate}[Independent systems] There exist independent physical systems.
\end{postulate}
\begin{postulate}[Informationally complete observable]\label{p:infocom} For each physical system
  there exists an informationally complete observable. 
\end{postulate}
\index{local observability principle} 
\begin{postulate}[Local observability principle]\label{p:locobs} For every composite system there exist
informationally complete observables made only of minimal local informationally complete observables.
\end{postulate}
\begin{postulate}[Informationally complete discriminating observable]\label{p:Bell} On every
  composite system made of two identical physical systems there exists a discriminating observable
  that gives a minimal informationally complete observable for one of the components, for some
  preparations of the other component. 
\end{postulate}
\begin{postulate}[Symmetric faithful state]\label{p:faith} For every composite system made of two identical
  physical systems there exist a symmetric joint state that is both dynamically and preparationally faithful. 
\end{postulate}
\medskip The General Axioms \ref{ga:1} and \ref{ga:2} entail a very rich series of notions, including
those used in the Postulates---\eg independent systems, observable, informationally complete
observable, etc.  In Sections \ref{s:states}-\ref{s:faithful}, starting from the two General Axioms,
I will introduce step by step such notions, starting from the pertaining definitions, and then
giving the logically related rules. For a discussion on the General Axioms the reader is addressed
to the publication \cite{darianoVax2005}, where also the generality of the definition of experiment
given in the General Axioms \ref{ga:1} is analyzed in some detail.
\section{Transformations, States, Independent systems}\label{s:states}
\par Performing a different experiment on the same object obviously
corresponds to the use of a different experimental apparatus or, at least, to a change of some settings of the
apparatus. Abstractly, this corresponds to change the set $\{\tA_j\}$ of possible transformations,
$\tA_j$, that the system can undergo.
\glossary{\Idx{transformations1}$\tA,\tB,\ldots,\tA_j,\tB_j,\ldots$ & transformations}
\index{transformation}  
Such change could actually mean
really changing the "dynamics" of the transformations, but it may simply mean changing only their
probabilities, or, just their labeling outcomes. Any such change actually corresponds to a change of
the experimental setup. Therefore, the set of all possible transformations $\{\tA_j\}$ will be
identified with the choice of experimental setting, \ie with the {\em experiment}
itself---or, equivalently, with the {\em action} of the experimenter: this will be formalized by the
following definition 
\begin{definition}[Actions/experiments and outcomes]\glossary{\Idx{actions}$\AA,\AB,\AC,\ldots$ & actions}
\index{action!definition}\index{outcome} An {\bf action} or {\bf experiment} on the
object system is given by the set $\AA\equiv\{\tA_j\}$ 
of possible transformations $\tA_j$ having overall unit probability,
with the apparatus signaling the {\bf outcome} $j$ labeling which
transformation actually occurred.  
\end{definition}
Thus the action/experiment is just a {\em complete} set of possible transformations that can
occur in an experiment.
As we can see now, in a general probabilistic framework the {\em action} $\AA$ is the "cause",
whereas the {\em outcome} $j$ labeling the transformation\index{cause and effect} $\tA_j$ that
actually occurred is the "effect". The {\em action} has to be regarded as the ``cause'', since it is the
option of the experimenter, and, as such, it should be viewed as deterministic (at least one
transformation $\tA_j\in\AA$  will occur with certainty), whereas the outcome $j$---\ie which
transformation $\tA_j$ occurs---is probabilistic.  
The special case of a deterministic transformation $\tA$ corresponds to a {\em singleton
  action/experiment} $\AA\equiv\{\tA\}$. 
\index{transformation!deterministic}
\medskip
\par In the following, wherever we consider a nondeterministic transformation $\tA$ by itself, we
always regard it in the context of an experiment, namely for any nondeterministic transformation
there always exists at least a complementary one $\tB$ such that the overall probability of
occurrence of one of them is always unit.  According to General Axiom \ref{ga:1} by definition the
knowledge of the state of a physical system allows us to predict the results of forthcoming possible
experiments on the system, or, more generally, on another system in the same physical situation.
Then, according to the General Axiom \ref{ga:2} a precise knowledge of the state of a system would
allow us to evaluate the probabilities of any possible transformation for any possible experiment.
It follows that the only possible definition of state is the following
\begin{definition}[States]\label{istate}\index{state(s)}
\glossary{\Idx{state1}$\omega,\zeta,\ldots$ & states}
\glossary{\Idx{state2}$\Omega,\Phi,\ldots$ & multipartite states}
 A  state $\omega$ for a physical
  system is a rule that provides the probability for any possible
  transformation, namely 
\begin{equation}
\omega:\textbf{state},\quad\omega(\tA):\text{probability that the
  transformation $\tA$ occurs}.
\end{equation}
\end{definition}
\medskip
We assume that the identical transformation $\tI$ occurs with probability one, namely 
\glossary{\Idx{transformations2}$\tI$ & identical transformation}
\begin{equation}
\omega(\tI)=1.\label{normcond}
\end{equation}
This corresponds to a kind of {\em interaction picture}, in which we do not consider the free
evolution of the system\index{intermediate/interaction picture}
(the scheme could be easily generalized to include a free evolution). Mathematically, a state will
be a map $\omega$ from the set of physical transformations to the
interval $[0,1]$, with the normalization condition
(\ref{normcond}). Moreover, for every action $\AA=\{\tA_j\}$ 
one has the normalization of probabilities \index{action!normalization condition}
\begin{equation}
\sum_{\tA_j\in\AA}\omega(\tA_j)=1
\end{equation}
for all states $\omega$ of the system. As already noticed in Ref. \cite{darianoVax2005}, in order to
include also non-disturbing experiments, one must conceive situations in which all states are left
invariant by each transformation.
\medskip
\par The fact that we necessarily work in the presence of partial knowledge about both object and
apparatus requires that the specification of the state and of the transformation could be given
incompletely/probabilistically, entailing a convex structure on states and an addition rule for
coexistent transformations.  The convex structure of states is given more precisely by the rule
\begin{grule}[Convex structure of states]\label{idim}\index{state(s)!convex structure}
The possible states of a physical system comprise a convex set: for any two states 
$\omega_1$ and $\omega_2$ we can consider the state $\omega$ which is
the {\em mixture} of $\omega_1$ and $\omega_2$, 
corresponding to have $\omega_1$ with probability $\lambda$ and
$\omega_2$ with probability $1-\lambda$. We will write
\begin{equation}
\omega=\lambda\omega_1+(1-\lambda)\omega_2,\quad 0\le\lambda\le 1,
\end{equation}
and the state $\omega$ will correspond to the following probability
rule for transformations $\tA$
\begin{equation}
\omega(\tA)=\lambda\omega_1(\tA)+(1-\lambda)\omega_2(\tA).
\end{equation}
\end{grule}
Generalization to more than two states is obtained by induction. In the
following the convex set of states will be denoted by $\Stset$. 
\glossary{\Idx{convex1}$\Stset$ & convex set of states}
We will call {\em pure} the states which are the extremal elements of
the convex set, namely which cannot be obtained as mixture of any two
states, and we will call {\em mixed} the non-extremal ones. As regards
transformations, the addition of coexistent transformations and
the convex structure will be considered in Rules \ref{g:addtrans} and \ref{r:convextrans}.
\bigskip
\begin{grule}[Transformations form a monoid]\label{isemigroup}
\index{transformation!semigroup of}\index{transformation!composition}
\index{transformation!monoid of}
The composition $\tA\circ\tB$ of two transformations $\tA$ and $\tB$
is itself a transformation. Consistency of compostion of transformations requires {\em
associativity}, namely\index{transformation!associativity} 
\begin{equation}
\tC\circ(\tB\circ\tA)=(\tC\circ\tB)\circ\tA.
\end{equation}
There exists the identical transformation $\tI$ which leaves the physical system invariant, and
which for every transformation $\tA$ satisfies the composition rule 
\begin{equation}
\tI\circ\tA=\tA\circ\tI=\tA.
\end{equation}
Therefore, transformations make a semigroup with identity, \ie a {\em monoid}.
\end{grule}
\begin{definition}[Independent systems and local experiments]\label{iindep}
  \index{independent systems}\index{independence}\index{experiment!local} We say that two physical
  systems are {\em independent} if on each system we can perform {\em local
experiments} that do not affect the other system for any joint state of the two systems. This can be
expressed synthetically with the commutativity of transformations of the local experiments, namely 
\begin{equation}
\tA^{(1)}\circ\tB^{(2)}=\tB^{(2)}\circ\tA^{(1)},
\end{equation}
where the label $n=1,2$ of the transformations denotes the system
undergoing the transformation.  
\end{definition}
In the following, when we have more than one independent system, we will denote local
transformations as ordered strings of transformations as follows
\begin{equation}\label{notlocal}
\tA,\tB,\tC,\ldots\doteq \tA^{(1)}\circ\tB^{(2)}\circ\tC^{(3)}\circ\ldots
\end{equation}
where the list of transformation on the left denotes the occurrence of local transformation $\tA$ on system 1,
$\tB$ on system 2, etc.
\glossary{\Idx{transformations3a}$(\tA,\tB,\tC,\ldots)$ & local transformations}
\glossary{\Idx{transformations3b}$\tA^{(1)}\circ\tB^{(2)}\circ\tC^{(3)}\circ\ldots$ & local transformations}
\section{Conditioned states and local states}
\begin{grule}[Bayes] When composing two transformations $\tA$ and $\tB$, the probability
$p(\tB|\tA)$ that $\tB$ occurs conditional on the previous occurrence of $\tA$ is given by the Bayes
rule\index{Bayes rule} 
\begin{equation}
p(\tB|\tA)=\frac{\omega(\tB\circ\tA)}{\omega(\tA)}.
\end{equation}
\end{grule}
The Bayes rule leads to the concept of {\em conditional
  state}:\index{state(s)!conditional}\index{conditional state}
\begin{definition}[Conditional state]\label{istatecond} The {\em conditional
state} $\omega_\tA$ gives the probability that a transformation
$\tB$ occurs on the physical system in the state $\omega$ after the
transformation $\tA$ has occurred, namely
\begin{equation}\label{condstate}
\omega_\tA(\tB)\doteq\frac{\omega(\tB\circ\tA)}{\omega(\tA)}.
\end{equation}
\glossary{\Idx{state3}$\omega_\tA$ & conditional state (state $\omega$ conditioned by the
  transformation $\tA$)}
\end{definition}
\par In the following we will make extensive use of the functional notation
\begin{equation}
\omega_\tA\doteq\frac{\omega(\cdot\circ\tA)}{\omega(\tA)},
\end{equation}
where the centered dot stands for the argument of the map. Therefore, the notion of conditional state describes
the most general {\em evolution}. 
\begin{definition}[Local state]\label{istateloc}
\index{state(s)!local}\index{local!state}
In the presence of many independent systems in a joint state $\Omega$, we define the {\bf
local state} $\Omega|_n$ of the $n$-th system the state that gives the probability for any local
transformation $\tA$ on the $n$-th system, with all other systems
untouched, namely
\begin{equation}
\Omega|_n(\tA)\doteq\Omega(\tI,\ldots,\tI,\underbrace{\tA}_{n\text{th}},\tI,\ldots).
\end{equation}
\glossary{\Idx{state4}$\Omega|_n$ & local state}
\end{definition}
For example, for two systems only, (which is equivalent to group
$n-1$ systems into a single one), we just write $\Omega|_1=\Omega(\cdot,\tI)$.
Notice that generally commutativity of local transformations (\ie Definition \ref{iindep}) does not 
imply that a transformation on system 2 does not affect the conditioned local state on system 1.
We also emphasize that acausality of local actions is not logically entailed by system independence 
 (for a discussion about acausality see Ref. \cite{darianoVax2005}).
\begin{remark}[Linearity of evolution]
  At this point it is worth noticing that the present definition of ``state'', which logically
  follows from the definition of experiment, leads to a {\em notion of evolution as
    state-conditioning}. In this way, each transformation acts linearly on the state space. In
  addition, since states are probability functionals on transformations, by dualism (equivalence
  classes of) transformations are linear functionals over the state space.
\end{remark}
\bigskip
\glossary{\Idx{state2}$\tilde\omega,\tilde\zeta,\ldots$ & weights}
\par For the following it is convenient to extend the notion of state to that of {\em weight}, \index{weight} 
namely nonnegative bounded functionals $\tilde\omega$ over the set of transformations with
$0\le\tilde\omega(\tA)\le\tilde\omega(\tI)<+\infty$ for all transformations $\tA$.  To each weight
$\tilde\omega$ it corresponds the properly normalized state
\begin{equation}
\omega=\frac{\tilde\omega}{\tilde\omega(\tI)}.
\end{equation}

Weights make the convex cone $\tilde\Stset$ which is generated by the convex set of states $\Stset$.
\glossary{\Idx{convex}$\tilde\Stset$ & convex cone $\tilde\Stset$ generated by the convex set of
  states}
\begin{definition}[Linear real space of generalized weights]
\index{generalized weights}
 We extend the notion of weight to that of negative
  weight, by taking differences. Such generalized weights span the affine linear space $\Wset$ of the
  convex cone of weights.
\end{definition}
\begin{remark} The transformations $\tA$ act as linear transformations over the space of weights as
  follows
\begin{equation}
\tA \tilde\omega=\tilde\omega(\tB\circ\tA).\label{wtransf} 
\end{equation}
\end{remark}

We are now in position to introduce the concept of {\em operation}.
\begin{definition}[Operation]\label{operation}\index{operation} 
To each transformation $\tA$ we can associate a linear map $\Op{\tA}:\;\Stset\longrightarrow\tilde\Stset$,
which sends a state $\omega$ into the unnormalized state $\tilde\omega_\tA\doteq
\Op{\tA}\omega\in\tilde\Stset$, defined
by the relation 
\begin{equation}
\Op{\tA}\omega\doteq\tilde\omega_\tA,\qquad\tilde\omega_\tA(\tB)=\omega(\tB\circ\tA).
\end{equation}
\end{definition}
Similarly to a state, the linear form $\tilde\omega_\tA\in\tilde\Stset$ for fixed $\tA$ maps from
the set of transformations to the interval $[0,1]$. It is not strictly a state only due to lack of
normalization, since $0<\tilde\omega_\tA(\tI)\le 1$. The operation $\operatorname{Op}$ 
gives the conditioned state through the state-reduction rule \index{state!state-reduction} \index{state-reduction}
\begin{equation}
\omega_\tA=\frac{\tilde\omega_\tA}{\omega(\tA)}
\equiv\frac{\Op{\tA}\omega}{\Op{\tA}\omega(\tI)}.
\end{equation}
\glossary{\Idx{operation}$\Op{\tA}$ & operation corresponding to the transformation $\tA$}
\bigskip
\section{Dynamical and informational structure}\label{s:transequivalence}
From the Bayes rule, or, equivalently, from the definition of
conditional state, we see that we can have the following complementary situations:
\begin{enumerate}
\item There are different transformations which produce the same state
  change, but generally occur with different probabilities;
\item There are different transformations which always occur with the
  same probability, but generally affect a different state change.
\end{enumerate}
The above observation leads us to the following definitions of
dynamical and informational equivalences of transformations.
\begin{definition}[Dynamical equivalence of transformations]\label{d:dyneq}
  \index{transformation!dynamical equivalence} \index{dynamical equivalence of transformations} Two
  transformations $\tA$ and $\tB$ are dynamically equivalent if $\omega_\tA=\omega_\tB$ for all
  possible states $\omega$ of the system.  We will denote the equivalence class containing the
  transformation $\tA$ as $[\tA]_{dyn}$.
\end{definition}
\medskip
\begin{definition}[Informational equivalence of transformations]
  \index{transformation!informational equivalence} \index{informational equivalence of
    transformations} Two transformations $\tA$ and $\tB$ are informationally equivalent if
  $\omega(\tA)=\omega(\tB)$ for all possible states $\omega$ of the system.  We will denote the
  equivalence class containing the transformation $\tA$ as $[\tA]$.
\end{definition}
\begin{definition}[Complete equivalence of transformations/experiments]\label{d:compleq}
\index{transformation!complete equivalence}
\index{experiment!complete equivalence}
\index{complete equivalence!of transformations}
\index{complete equivalence!of experiments}
Two transformations/experiments are completely equivalent iff they are both
dynamically and informationally equivalent.
\end{definition}
Notice that even though two transformations are completely equivalent,
in principle they can still be different experimentally, in the sense
that they are achieved with different apparatus. However, we emphasize that outcomes
in different experiments corresponding to equivalent transformations always provide
the same information on the state of  the object, and, moreover, the
corresponding transformations of the state are the same.
The concept of dynamical equivalence of transformations leads one to
introduce a convex structure also for transformations. We first need
the notion of informational compatibility.
\index{transformation!coexistence}
\index{transformation!informational compatibility}
\index{informational compatibility of transformations}
\index{coexistence of transformations}
\begin{definition}[Informational compatibility or coexistence] We say that
two transformations $\tA$ and $\tB$ are {\em coexistent} or {\em
informationally compatible} if one has 
\begin{equation}
\omega(\tA)+\omega(\tB)\le 1,\quad\forall\omega\in\Stset,\label{compatible}
\end{equation}
\end{definition}
The fact that two transformations are coexistent means that, in principle, they can occur in the
same experiment, namely there  
exists at least an action containing both of them. We have named the
present kind of compatibility "informational" since it is actually
defined on the informational equivalence classes of transformations.
\par We are now in position to define the "addition" of coexistent transformations.
\begin{grule}[Addition of coexistent transformations]\label{g:addtrans}
\index{transformation!addition} For any two
  coexistent transformations $\tA$ and $\tB$  we define the
  transformation $\tS=\tA_1+\tA_2$ as the transformation corresponding
  to the event $e=\{1,2\}$, namely the apparatus signals that either
  $\tA_1$ or $\tA_2$ occurred, but does not specify which one.
By definition, one has the distributivity rule 
\begin{equation}\label{r:sum1}
\forall\omega\in\Stset\qquad\omega(\tA_1+\tA_2)=\omega(\tA_1)+\omega(\tA_2),
\end{equation}
whereas the state conditioning is given by
\begin{equation}\label{r:sum2}
\forall\omega\in\Stset\qquad
\omega_{\tA_1+\tA_2}=\frac{\omega(\tA_1)}{\omega(\tA_1+\tA_2)}
\omega_{\tA_1}+\frac{\omega(\tA_2)}{{\omega(\tA_1+\tA_2)}}\omega_{\tA_2}.
\end{equation}
\glossary{\Idx{transformations5}$\tA+\tB$ & addition of compatible transformations}
\end{grule}
Notice that the two rules in Eqs.  (\ref{r:sum1}) and (\ref{r:sum2}) completely specify the 
transformation $\tA_1+\tA_2$, both informationally and dynamically. Eq. (\ref{r:sum2}) can be more
easily restated in terms of operations as follows:
\begin{equation}\label{r:addtrans}
\forall\omega\in\Stset\qquad
\Op{\tA_1+\tA_2}\omega=\Op{\tA_1}\omega+\Op{\tA_2}\omega.
\end{equation}
Addition of compatible transformations is the core of the description
of partial knowledge on the experimental apparatus. Notice also that
the same notion of coexistence can be extended to "propensities" as well
(see Definition \ref{d:propensity}). 
\begin{grule}[Multiplication of a transformation by a scalar]\label{g:scalmult}
\index{transformation!multiplication by a scalar}
For each transformation $\tA$ the transformation $\lambda\tA$ for
$0\le\lambda\le 1$ is defined as the transformation which is 
dynamically equivalent to $\tA$, but which occurs with probability
$\omega(\lambda\tA)=\lambda\omega(\tA)$.   
\end{grule}
Notice that according to Definition \ref{d:compleq} two transformations are completely characterized
operationally by the informational and dynamical equivalence classes to which they belong, whence
Rule \ref{g:scalmult} is well posed.
\begin{remark}[Algebra of generalized transformations]\label{rem:transf}
  Using Eqs. (\ref{r:sum1}) and (\ref{r:addtrans}) one can extend the addition of coexistent
transformations to generic linear combinations: the {\em generalized transformations}.
\index{generalized transformations} The generalized transformations constitute a real vector space,
which is the affine space of the convex space $\Trnset$. Composition of transformations can be
extended via linearity to generalized transformations, making their space a real algebra $\aA$, the
{\em algebra of generalized transformations}. Notice that every generalized transformation belongs
to the dynamical equivalence class of a physical transformation, since the conditioned state is
always defined. 
\end{remark}
\bigskip
It is now natural to introduce a norm over transformations as follows.
\begin{theorem}[Norm for transformations]\label{t:transnorm} The following quantity
\index{transformation!norm}
\glossary{\Idx{transformations4}$\n{\tA}$ & norm of transformation}
\begin{equation}
\n{\tA}=\sup_{\omega\in\Stset}\omega(\tA),\label{norm}
\end{equation}
is a norm on the set of transformations. In terms of such norm all transformations are contractions.
\index{transformation!contraction}
\end{theorem}
\Proof We remind the axioms of norm:
i) Sub-additivity $\n{\tA+\tB}\le\n{\tA}+\n{\tB}$; ii) Multiplication by scalar
$\n{\lambda\tA}=\lambda\n{\tA}$; iii) $\n{\tA}=0$ implies $\tA=0$.
The quantity in Eq. (\ref{norm}) satisfy the sub-additivity relation i), since
\begin{equation}
\n{\tA+\tB}=\sup_{\omega\in\Stset}[\omega(\tA)+\omega(\tB)]\le
\sup_{\omega\in\Stset}\omega(\tA)+\sup_{\omega'\in\Stset}\omega'(\tB)=
\n{\tA}+\n{\tB}.
\end{equation}
Moreover, it obviously satisfies axiom ii). Finally, axiom iii) corresponds to identify all
transformations that never occur (occur with zero probability) with the zero transformation $\tA=0$.
It is also clear that, by definition, for each transformation $\tA$ one
has $\n{\tA}\le1$, namely transformations are contractions.\qed
We remind that the multiplication of a transformation $\tA$ by a scalar is still a transformation
only for scalar $0\le\lambda\le\n{\tA}^{-1}$.
\begin{theorem}\label{t:submult}
The norm in Eq. (\ref{norm}) satisfies the following inequality
\begin{itemize}
\item[iv)]\label{nAB} $\n{\tB\circ\tA}\le\n{\tB}\n{\tA}.$
\end{itemize}
\end{theorem}
\Proof Using the definition of conditional state in Eq. (\ref{condstate}) we have
\begin{equation}
\n{\tB\circ\tA}=\sup_{\omega\in\Stset}\omega(\tB\circ\tA)=
\sup_{\omega\in\Stset}\omega_\tA(\tB)\omega(\tA)\le 
\sup_{\omega\in\Stset}\omega_\tA(\tB)\sup_{\zeta\in\Stset}\zeta(\tA) \le 
\sup_{\omega\in\Stset}\omega(\tB)\sup_{\zeta\in\Stset}\zeta(\tA)=\n{\tB}\n{\tA}.
\end{equation}
\qed
\par The linear space of generalized weights $\Wset$ can also be equipped with a norm. For this we
need to introduce the following notion of experimentally sufficient set of transformations.
\begin{theorem}[Norm over generalized weights] The following is a norm over  generalized weights
\begin{equation}
\n{\tilde\omega}=\sup_{\tA\in\Trnset}|\tilde\omega(\tA)|.\label{normstate}
\end{equation}
\end{theorem}
\Proof The quantity in Eq. (\ref{normstate}) satisfies the sub-additivity relation 
$\n{\tilde\omega+\tilde\zeta}\le\n{\tilde\omega}+\n{\tilde\zeta}$, since 
\begin{equation}
\n{\tilde\omega+\tilde\zeta}=\sup_{\tA\in\Trnset}[|\tilde\omega(\tA)+\tilde\zeta(\tA)|]\le
\sup_{\tA\in\Trnset}[|\tilde\omega(\tA)|+|\tilde\zeta(\tA)|]\le
\sup_{\tA\in\Trnset}|\tilde\omega(\tA)|+\sup_{\tA\in\Trnset}|\tilde\zeta(\tA)]|=
\n{\tilde\omega}+\n{\tilde\zeta}.
\end{equation}
Moreover, it obviously satisfies the identity
\begin{equation}
\n{\lambda\tilde\omega}=|\lambda|\n{\omega}.
\end{equation}
Finally, $\n{\tilde\omega}=0$ implies that $\tilde\omega=0$, since either $\tilde\omega$ is a
positive linear form, \ie it is proportional to a true state, whence at least $\tilde\omega(\tI)>0$,
or $\tilde\omega$ is the difference of two positive linear forms, whence the two corresponding
states must be equal by definition, since their probability rules are equal, which means that,
again, $\tilde\omega=0$.
\qed
\begin{remark}[Banach space of  generalized weights]\label{r:banw}
Closure with respect to the norm (\ref{normstate}) makes the real vector space of generalized
weights $\Wset$  a Banach space, which we will name the {\em Banach space of generalized weights}. 
The norm closure correspond to assume the possibility of preparing states with probabilities close
to that of a given one, with the approximability criterion defined by the norm.
\end{remark}
\begin{remark}[Norms, approximability criteria, and norm closure]\label{r:closure} Norms defined as in
  Eq. (\ref{norm}) or Eq. (\ref{normstate}) (see also other norms in the following) operationally
  correspond to approximability criteria.  The norm closure is not operationally required, but, as
  any other kind of extension, it is mathematically convenient. Therefore, in the following we
  should remind that if norm closure is not operationally assumed in terms of a separate postulate
  (clearly not of operational nature), then the Banach space element---\eg the limit of a Cauchy
  sequence---does not necessarily correspond to a physically achievable quantity.
\end{remark}
\bigskip
\par In terms of the norm (\ref{norm}) for transformations one can equivalently define coexistence
(informational compatibility) using the following corollary
\begin{corollary} Two transformations $\tA$ and $\tB$ are
  coexistent iff $\tA+\tB$ is a contraction.
\end{corollary}
\Proof If the two transformations are coexistent, then
from Eqs. (\ref{compatible}) and (\ref{norm}) one has that
$\n{\tA+\tB}\le 1$. On the other hand, if $\n{\tA+\tB}\le 1$, this
means that Eq. (\ref{norm}) is satisfied for all states, namely the
transformations are coexistent.\qed 
\begin{corollary} The transformations $\lambda\tA$ and
  $(1-\lambda)\tB$ are compatible for any couple of transformations
  $\tA$ and $\tB$. 
\end{corollary}
\Proof Clearly
$\n{\lambda\tA+(1-\lambda)\tB}\le\lambda\n{\tA}+(1-\lambda)\n{\tB}\le 1$.\qed 
\medskip
\par The last corollary implies the rule
\begin{grule}[Convex structure of transformations]\label{r:convextrans}
\index{transformation!convex structure}
Transformations form a convex set, namely for any two 
transformations $\tA_1$ and $\tA_2$ we can consider the transformation
$\tA$ which is the {\em mixture} of $\tA_1$ and $\tA_2$ with
probabilities $\lambda$ and $1-\lambda$. Formally, we write
\begin{equation}
\tA=\lambda\tA_1+(1-\lambda)\tA_2,\quad 0\le\lambda\le 1,\label{ala}
\end{equation}
with the following meaning: the transformation $\tA$ is itself a
probabilistic transformation, occurring with overall probability
\begin{equation}
\omega(\tA)=\lambda\omega(\tA_1)+(1-\lambda)\omega(\tA_2),
\end{equation}
meaning that when the transformation $\tA$ occurred we
know that the transformation dynamically was either $\tA_1$ with 
(conditioned) probability $\lambda$ or $\tA_2$ with
probability $(1-\lambda)$.
\end{grule}
We have seen that the transformations form a convex set, more 
specifically, a spherically truncated convex cone, namely we can
always add transformations or multiply a transformation by a positive scalar if the result is a contraction. In
the following we will denote the spherically truncated convex cone of transformations 
as $\Trnset$. 
\glossary{\Idx{convex3}$\Trnset$ & truncated convex cone of transformations}
\medskip
\begin{remark} The norm (\ref{norm}) can be extended to the whole algebra $\aA$ of generalized
  transformations as follows
\begin{equation}\label{seminorm}
\n{\tA}=\sup_{\omega\in\Stset}|\omega(\tA)|.
\end{equation}
It is then easy to check the axioms i), and ii) of norm. However, axiom iii) does not hold anymore,
since one has $\n{\tC}=0$ for $\tC=\tA-\tB$ with $\tA$ and $\tB$ informationally
equivalent transformations. Therefore, the norm extension in Eq. (\ref{seminorm}) is  only a {\em
  seminorm}. \index{seminorm on generalized transformations}  Also the bound (\ref{nAB}) is not
meaningful for the extension, since for the same $\tA$ above one would have $\omega(\tA)=0$. We
conclude that we cannot introduce the structure of Banach algebra over $\aA$. A Banach space
structure can, however, be introduced for the affine space of propensities (see the following).
\end{remark}
\section{Propensities}
Informational equivalence allows one to define equivalence classes of transformations, which we may
want to call {\em propensities}, since they give the occurrence probability of a transformation for
each state, \ie its ``disposition'' to occur.  
\begin{definition}[Propensities]\label{d:propensity}
\index{propensity} We call {\bf propensity} an
  informational equivalence class of transformations.  
\end{definition}
It is easy to see that the present notion of propensity corresponds
closely to the notion of "effect" introduced by Ludwig
\cite{Ludwig-axI}. However, we prefer to keep a separate word, since
the "effect" has been identified with a quantum mechanical notion and
a precise mathematical object (\ie a positive contraction).
\glossary{\Idx{propensity1}$\cA,\cB,\ldots,$ & propensities}
\glossary{\Idx{propensity2}$[\tA]$ & propensity containing the transformation $\tA$}
In the following we will denote propensities with underlined symbols
as $\cA$, $\cB$, etc., and we will use the notation $[\tA]$ for the
propensity containing the transformation $\tA$, and also write
$\tA_0\in[\tA]$ to say that $\tA_0$ is informationally equivalent to
$\tA$. Thus, by definition one has $\omega(\tA)\equiv\omega([\tA])$, and one can legitimately
write $\omega(\cA)$. Similarly, one has $\tilde\omega_\tA(\tB)\equiv
\tilde\omega_\tA([\tB])$ which implies that  $\omega(\tB\circ\tA)=\omega([\tB]\circ\tA)$ which gives
the chaining rule
\begin{equation}
[\tB]\circ\tA\subseteq[\tB\circ\tA].
\end{equation}
One also has the locality rule
\begin{equation}
[(\tA,\tB)]\supseteq([\tA],[\tB]),
\end{equation}
where we used notation (\ref{notlocal}).
It is clear that $\lambda\tA$ and $\lambda\tB$ belong to the
same equivalence class iff $\tA$ and $\tB$ are informationally
equivalent. This means that also for propensities multiplication
by a scalar can be defined as $\lambda[\tA]=[\lambda\tA]$. Moreover,
since for $\tA_0\in[\tA]$ and $\tB_0\in[\tB]$ one has 
$\tA_0+\tB_0\in[\tA+\tB]$, we can define addition of propensities as
\index{propensity!addition}
$[\tA]+[\tB]=[\tA+\tB]$ for any choice of representatives $\tA$ and
$\tB$ of the two added propensities. Also, since all transformations
of the same equivalence class have the same norm, we can extend
the definition (\ref{norm}) to propensities as $\n{[\tA]}=\n{\tA}$ for
any representative $\tA$ of the class. It is easy to check
sub-additivity on classes, which implies that it is indeed a norm. In
fact, one has 
\begin{equation}
\n{[\tA]+[\tB]}=\n{\tA+\tB}\le\n{\tA}+\n{\tB}=\n{[\tA]}+\n{[\tB]}.
\end{equation}
Therefore, it follows that also propensities form a spherically
truncated convex cone, which we will denote by $\Cntset$.
\glossary{\Idx{convex5}$\Cntset$ & truncated convex cone of propensities}
\begin{remark}[Banach space of generalized propensities] The norm for propensities can be extended to the
  embedding affine space of $\Cntset$. One can see that in this case all axioms of norm hold, and
  one can construct a Banach space, with the norm-closure corresponding to an approximation
  criterion for propensities (see also Remark \ref{r:closure}).
\end{remark}
\medskip
\glossary{\Idx{propensity3}$l$ & propensity}
\begin{remark}[Duality between the convex sets of states and of propensities]
From the Definition \ref{istate} of state it follows that the
convex set of states $\Stset$ and the convex sets of propensities
$\Cntset$ are dual each other, and the latter can be regarded as the
set of positive linear contractions over the set of states, namely the
set of positive functionals $l$ on $\Stset$ with unit upper bound, and
with the functional $l_{[\tA]}$ corresponding to the propensity $[\tA]$
being defined as
\glossary{\Idx{propensity4}$l_{[\tA]}$ & propensity containing the transformation $\tA$}
\begin{equation}
l_{[\tA]}(\omega)\doteq\omega(\tA).
\end{equation}
In the following we will often identify propensities with their
corresponding functionals, and denote them by lowercase letters
$a,b,c,\ldots$, or $l_1,l_2,\ldots$. Finally, notice that the notion of
coexistence (informational compatibility) extends naturally to
propensities. 
\end{remark}
\begin{definition}[Observable]\index{observable} We call 
  observable a set of propensities $\AL=\{l_i\}$ which is informationally equivalent to an action
  $\AL\in\underline{\AA}$, namely such that there exists an action $\AA=\{\tA_j\}$ for which one
  has $l_i\in\underline{\tA_j}$.
\end{definition}
Clearly, the generalized observable is normalized to the constant unit functional, \ie $\sum_il_i=1$.
\begin{definition}[Informationally complete observable] An observable
  $\AL=\{l_i\}$ is informationally complete if each propensity can be written as a linear combination of
  the of elements of $\AL$, namely for each propensity $l$ there exist coefficients $c_i(l)$ such that
\begin{equation}
l=\sum_ic_i(l)l_i.
\end{equation}
We call the informationally complete observable {\em minimal} \index{informationally
  complete observable!minimal} when its propensities are linearly independent.
\end{definition}
Clearly, using an informationally complete observable one can reconstruct any state
$\omega$ from just the probabilities $l_i(\omega)$, since one has
\begin{equation}
\omega(\tA)=\sum_ic_i(l_{\underline{\tA}})l_i(\omega).
\end{equation}
Based on the notion of informationally complete observable, we can introduce the following one
\begin{definition}[Experimentally sufficient set of transformations] \index{experimentally
    sufficient set of transformations} \index{transformations!experimentally sufficient set of}  We
  call a set of transformations $\trnset$ {\em experimentally sufficient} if it has a subset that is in
  correspondence with an informationally complete observable. 
\end{definition}
Using the above notion we can introduce a norm $\n{\cdot}_\trnset$ for generalized weights,
generalizing the norm given in Eq. (\ref{normstate}), by taking the supremum over $\trnset$ instead
of $\Trnset$. The fact that the set of transformations is experimentally sufficient guarantees that 
$\n{\tilde\omega}_\trnset=0$ implies that $\tilde\omega=0$. The restriction to a set $\trnset$ of
transformations may be operationally motivated. An analogous restriction may be considered
for the norm of generalized transformations, by restricting the set of states $\Stset$.
\begin{definition}[Predictability and resolution]\label{def:res} 
\index{transformation!predictable}\index{predictable!transformation}
\index{propensity!predictable}\index{predictable!propensity}
We will call a transformation $\tA$---and likewise its
propensity---{\em predictable} if there exists a state for which
$\tA$ occurs with certainty and some other state for which it never
occurs. The transformation (propensity) will be also called {\em
resolved} if the state for which it occurs with certainty is
unique---whence pure.
An action will be called {\em predictable} when it is made only
of predictable transformations, and {\em resolved} when all
transformations are resolved.
\end{definition}
The present notion of predictability for propensity corresponds to that of
"decision effects" of Ludwig \cite{Ludwig-axI}. For a predictable
transformation $\tA$ one has $\n{\tA}=1$. Notice that 
a predictable transformation is not deterministic, and it can
generally occur with nonunit probability on some state $\omega$. 
Predictable propensities $\tA$ correspond to affine functions
$f_\tA$ on the state space $\Stset$ with $0\le f_\tA\le 1$ achieving
both bounds. Their set will be denoted by $\Prdset$.

\begin{definition}[Perfectly discriminable set of states] We call a set of states $\{\omega_n\}_{n=1,N}$
  {\em perfectly discriminable} if there exists an action $\AA=\{\tA_j\}_{j=1,N}$ with transformations
  $\tA_j\in l_j$ corresponding to a set of  predictable propensities $\{l_n\}_{n=1,N}$ satisfying the 
  relation \index{state(s)!perfectly discriminable}
\begin{equation}
l_n(\omega_m)=\delta_{nm}.
\end{equation}
\end{definition}
\begin{definition}[Informational dimensionality] We call {\em informational dimension}
\index{dimension!informational} of the convex set of states $\Stset$, denoted by $\idim{\Stset}$,
the maximal cardinality of perfectly discriminable set of states in $\Stset$.
\glossary{\Idx{dimension3}$\idim{\Stset}$ & informational dimension of the convex set of states $\Stset$}
\end{definition}
\begin{definition}[Discriminating observable]
  \index{observable!discriminating}\index{propensity!observable} An observable
  $\AL=\{l_j\}$ is {\em discriminating for} $\Stset$ when it discriminates a set of
  states with cardinality equal to the informational dimension $\idim{\Stset}$ of $\Stset$.
\end{definition}
\bigskip
\section{Faithful state}\label{s:faithful}
\begin{definition}[Dynamically faithful state] 
\index{state!dynamically faithful}\index{dynamically faithful state}
We say that a state $\Phi$ of a composite system is {\em dynamically faithful} for the $n$th component
system when acting on it with a transformation $\tA$ results in an (unnormalized) conditional state
that is in  one-to-one correspondence with the dynamical equivalence class $[\tA]_{dyn}$ of $\tA$, namely the
following map is one-to-one:  
\glossary{\Idx{dyn}$[\tA]_{dyn}$ & dynamical equivalence class of transformations $\tA$}
\begin{equation}
\tilde\Phi_{\tI,\ldots,\tI,\tA,\tI,\ldots} \leftrightarrow [\tA]_{dyn},
\end{equation}
where in the above equation the transformation $\tA$ acts locally only on the $n$th component system.
\end{definition}
\begin{figure}[hbt]
    \setlength{\unitlength}{800sp}
    \begin{picture}(8745,3219)(931,-3565)
      {\thicklines \put(5401,-1261){\oval(1756,1756)}}
      {\put(1801,-1261){\line(1, 0){2700}}}
      {\put(6301,-1261){\vector(1, 0){3300}}}
      {\put(1801,-3361){\vector(1, 0){7800}}}
      \put(2026,-2611){\makebox(0,0)[b]{$\Phi$}}
      \put(5401,-1486){\makebox(0,0)[b]{$\tA$}}
      \put(9676,-2811){\makebox(0,0)[b]{$\Phi_{\tA,\tI}$}}
    \end{picture}
\caption{Illustration of the notion of dynamically faithful state: the conditioned state
  $\Phi_{\tA,\tI}$ is in one-to-one correspondence with the dynamical equivalence class of the
  transformation $\tA$.}
  \end{figure}
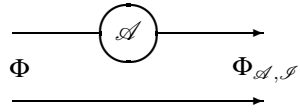
\medskip
\par In the following for simplicity we restrict attention to two component systems,
and take the first one for the $n$th. Using the definition \ref{istatecond} of conditional state, we
see that the state $\Phi$ is dynamically faithful when the map $\Phi(\cdot\circ[\tA,\tI]_{dyn})$ is 
invertible over the set of dynamical equivalence classes of transformations, namely when
\begin{equation}
\forall\tB,\;\Phi(\tB\circ(\tA_1,\tI))=\Phi(\tB\circ(\tA_2,\tI))\quad\Longleftrightarrow\quad\tA_1\in[\tA_2]_{dyn}.
\end{equation}
\begin{definition}[Preparationally faithful state]\label{prepfaith}
  \index{state!preparationally faithful}\index{preparationally faithful state} We will call a state
  $\Phi$ of a bipartite system {\em preparationally faithful} if all local states of one component
  can be achieved by a suitable local transformation of the other, namely for every state $\omega$
  of the first party there exists a local transformation $\tT_\omega$ of the other party for which
  the conditioned local state coincides with $\omega$, namely
\begin{equation}
\forall\omega\in\Stset\qquad\exists \tT_\omega:\qquad
\omega=\Phi_{\tT_\omega,\tI}|_2\doteq \frac{\Phi(\tT_\omega,\cdot)}{\Phi(\tT_\omega,\tI)}.
\end{equation}
\end{definition}
In Postulate \ref{p:faith} we also use the notion of {\em symmetric} joint state, defined as follows.
\begin{definition}[Symmetric joint state of two identical systems] We call a joint state of two
  identical systems {\em symmetric} if for a particular choice of local informationally complete
  measurements one has
\begin{equation}
\Phi(\cX_i,\cX_j)=\Phi(\cX_j,\cX_i),\quad\forall i,j.
\end{equation}
\end{definition}
We clearly have
\begin{equation}
\Phi(\cA,\cB)=\Phi(\cB,\cA),
\end{equation}
for any couple of propensities $\cA$ and $\cB$. Therefore, the choice of the local informationally complete
measurement is irrelevant. Moreover, for a symmetric faithful state we have 
\begin{equation}\label{faith4}
\Phi|_1(\tA)=\Phi|_1(\tA')=\Phi|_2(\tA)=\Phi|_2(\tA'),
\end{equation}
and for a symmetric preparationally faithful state we have
\begin{equation}
\omega=\Phi_{\tT_\omega,\tI}|_2=\Phi_{\tI,\tT_\omega}|_1.
\end{equation}
\section{The Bloch representation}\label{s:Bloch}
In this section we introduce an affine-space representation based on the existence of a minimal
informational complete observable. Such representation generalizes the popular Bloch representation
used in Quantum Mechanics. 
\par Let's fix a minimal informationally complete observable, denoted by $\{n_j\}$, 
in terms of which we can expand (in a unique way) any propensity as follows
\begin{equation}
l_{\cA}=\sum_jm_j(\cA)n_j.
\end{equation}
It is convenient to replace one element of the informationally complete observable $\{n_j\}$ with
the {\em normalization functional} $n_0$ defined as
\begin{equation}
n_0(\tilde\omega)=\tilde\omega(\cI),\qquad\forall \tilde\omega\in\tilde\Stset,
\end{equation}
[$n_0(\omega)=1$ for normalized states $\omega$]. We will then use the Minkowskian notation
\begin{equation}
n\doteq(n_0,\vec n),\;m\doteq(m_0,\vec m),
\qquad mn\doteq \sum_jm_jn_j=\vec m\cdot\vec n+m_0n_0.
\end{equation}
In the following we will also denote $q\doteq m_0$. Therefore, for any propensity $\cA$, we will write
\begin{equation}
l_{\cA}(\omega)=m(\cA)n(\omega)\equiv\vec m(\cA)\cdot\vec n(\omega)+q(\cA).
\end{equation}
Clearly one can extend the convex set of propensities $\Cntset$ to the complexification
$\Cmplx\Cntset$ of the underlying affine space, by keeping the coefficients $m_j$ of the expansion
as complex, namely a generic element $l\in\Cmplx\Cntset$ will be given by 
\begin{equation}
l=\sum_jm_jn_j,\quad m_j\in\Cmplx.
\end{equation}
Notice that $\vec n(\omega)$ gives a complete description of the state $\omega$, since for any
transformation $\tA$ one can write
\begin{equation}
\omega(\tA)=\vec m(\cA)\cdot\vec n(\omega)+q(\cA).
\end{equation}
On the other hand, by denoting with $\cX_j$ and $l_j$ the propensity such that 
$[\vec m(\cX_j)]_l=\delta_{jl}$ we have
\begin{equation}
\vec n_j(\omega)=l_{\cX_j}(\omega)\doteq l_j(\omega).
\end{equation}
Notice that $\cX_0\equiv\cI$. We will call $\vec n(\omega)$ the Bloch vector representing the state
$\omega$. The Bloch representation is {\em faithful} (\ie one-to-one), since the informationally
complete observable $\{l_j\}$ is minimal, namely the functionals $l_j$ are linearly independent. We
also emphasize that the representation is trivially extended to generalized weights, transformations
and propensities.
\par We now recover the linear transformation describing conditioning, given in terms of the {\em
  operation}, which we remind is given in terms of the unnormalized state
$\Op{\tA}\omega\equiv\tilde\omega_\tA$ defined as follows
\begin{equation}
\Op{\tA}\omega(\tB)\equiv\tilde\omega_\tA(\tB)=\omega(\tB\circ\tA)=\omega(\cB\circ\tA)
\equiv l_{\cB}(\tilde\omega_{\tA}).
\end{equation}
From linearity of transformations (see Eq. (\ref{r:addtrans}) and Remark \ref{rem:transf}), upon
introducing a matrix $\{M_{jl}(\tA)\}$, one can write
\begin{equation}\label{matrixM} 
\omega(\cX_j\circ\tA)=\sum_l M_{jl}(\tA)l_l(\omega)+ M_{j0}(\tA),
\end{equation}
and, in particular, 
\begin{equation}
\omega(\tX_0\circ\tA)\equiv\omega(\tA)=\sum_l M_{0l}(\tA)n_l(\omega)
\equiv \vec m(\cA)\cdot\vec n(\omega)+q(\cA),
\end{equation}
from which we derive the identities
\begin{equation}
M_{0l}(\tA)\equiv [\vec m(\cA)]_l,\qquad M_{00}(\tA)\equiv q(\cA).
\end{equation}
The real matrices $M_{jl}(\tA)$ are a {\em representation} of the real algebra of transformations $\aA$. 
The first row of the matrix is a representation of the propensity $\cA$ (see Fig. \ref{f:matA}).

In the Bloch-vector notation, one has
\begin{equation}
\vec n_j(\tilde\omega_{\tA})=l_{\cX_j}(\tilde\omega_{\tA})=\omega(\tX_j\circ\tA),\qquad
n_0(\tilde\omega_{\tA})=l_{\cX_0}(\tilde\omega_{\tA})=\omega(\tA).
\end{equation}

\begin{equation}\label{transfBloch}
\vec n(\tilde\omega_{\tA})=\vec M(\tA)\vec n(\omega)+\vec k(\tA),\qquad
\vec k_j(\tA)\doteq q(\tX_j\circ\tA),\qquad
n_0(\tilde\omega_{\tA})=\vec m(\cA)\cdot\vec n(\omega)+q(\cA),
\end{equation}

\begin{equation}
\tilde\omega_{\tA}(\tB)=\vec m(\tB)\cdot\vec n(\tilde\omega_{\tA})+q(\tB)n_0(\tilde\omega_{\tA})
\end{equation}

The matrix representation of the transformation is synthesized in Fig. \ref{f:matA}.
\begin{figure}[h]
$
M_{ij}(\tA)=\begin{pmatrix}
\fbox{\parbox[t][12mm][c]{12mm}{\begin{center}$q(\cA)$\end{center}}}
 &\!\!\! \!\fbox{\parbox[t][12mm][c]{22mm}{\begin{center}$\vec m(\cA)$\end{center}}} \\
\fbox{\parbox[t][22mm][c]{12mm}{\begin{center}$\vec k(\tA)$\end{center}}}
& \!\!\! \!\fbox{\parbox[t][22mm][c]{22mm}{\begin{center}$\vec M(\tA)$\end{center}}}
\end{pmatrix}
$
\caption{\small Matrix representation of the real algebra of transformations $\aA$. The first row represents the
  propensity $\cA$ of the transformation $\tA$. It gives the transformation of the zero-component of
  the Bloch vector $n_0(\tilde\omega_\tA)\equiv\omega(\tA)=\vec m(\cA)\cdot\vec n(\omega)+q(\cA)$,
  namely the probability of the transformation. The following rows represent the affine
  transformation of the Bloch vector $\vec n(\omega)$ corresponding to the quantum operation
  $\Op{\tA}$, the  first column giving the translation $\vec k(\tA)$, and the remaining square
  matrix $\vec M(\tA)$ the linear part. Overall, the Bloch vector of the state $\omega$ is transformed
  as $\vec n(\Op{\tA}\omega)=\vec M\vec n(\omega)+\vec k(\tA)$.\label{f:matA}}
\end{figure}
Since the Bloch representation is faithful, then the dimension of the affine space of the Bloch
vector $\vec n(\omega)$ is just the affine dimension $\adm(\Stset)$ of the convex set of states
$\Stset$.
\par Therefore, summarizing we have the following conditioning transformation
\begin{equation}
\vec n(\omega)\longrightarrow\vec n(\omega_\tA)=\frac{\vec M(\tA)\vec n(\omega)+\vec k(\tA)}{\vec m(\cA)\cdot\vec n(\omega)+q(\cA)},
\end{equation}
with the transformation occurring with probability given by
\begin{equation}
p(\tA;\omega)=\vec m(\cA)\cdot\vec n(\omega)+q(\cA).
\end{equation}
\par Using a joint local informationally complete observable, we can build a Bloch representation of
joint states and of transformations of the composed system. We introduce the dual tensor notation
$\vec n\odot\vec n$ with the following meaning
\begin{equation}
( n\odot n)_{ij}(\Phi)\equiv n_i\odot n_j(\Phi)\doteq l_{\cX_i,\cX_j}(\Phi),\quad i,j=0,1,\ldots
\end{equation}
and with the matrix composition rule
\begin{equation}
( M(\tA)\odot M(\tB))( n\odot n)(\Phi)=( M(\tA) n\odot
 M(\tA) n)(\Phi),
\end{equation}
corresponding to the probability rule
\begin{equation}
\Phi(\cX_i\circ\tA,\cX_j\circ\tB)=(\vec M(\tA)\vec n\odot\vec M(\tB)\vec n)_{ij}(\Phi)
\end{equation}
which follows from Eq. (\ref{matrixM}) along with the conditioning rule and the notion of local
state. For example, more explicitely for $i,j=1,2,\ldots$, one has
\begin{equation}
\begin{split}
\Phi(\cX_i\circ\tA,\cX_j\circ\tB)=&(\vec M(\tA)\vec n\odot\vec M(\tB)\vec n)_{ij}(\Phi)
+(\vec k(\tA))n_0\odot\vec M(\tB)\vec n)_{ij}(\Phi)\\
+&(\vec M(\tA))\vec n\odot\vec k(\tB) n_0)_{ij}(\Phi)+\vec k_i(\tA)\vec k_j(\tB)
\end{split}
\end{equation}
where we used the identity $(n_0\odot n_0)(\Phi)=1$.
It is easy to see that the representation of the local states $\Omega|_1=\Omega(\cdot,\tI)$
and  $\Omega|_2=\Omega(\tI,\cdot)$ are simply given by
\begin{equation}
n(\Omega|_1)=(n\odot n_0)(\Omega),\quad n(\Omega|_2)=(n_0\odot n)(\Omega).\label{locmat}
\end{equation}
\section{Operational adjoint and real Hilbert space structure}\label{s:Hilbert}
In this section we will see how it is possible to define operationally a real adjoint map (\ie a transposition)
using a symmetric faithful state, and how using such adjoint one can introduce a Hilbert
space structure via two different constructions: the Mackey-Kakutani and the Gelfand-Naimark-Segal
constructions. 
\subsection{Twin involution}
\par We now define the {\em twin} involution over transformations.
\begin{definition}
For a {\em faithful} bipartite state $\Phi$, the {\em twin} $\tA'$ of the transformation $\tA$
is that transformation which when applied to the second component system gives the same conditioned state
and with the same probability than the transformation $\tA$ operating on the first system. In
equations, one has
\begin{equation}\label{twinid}
\tilde\Phi_{\tA,\tI}=\tilde\Phi_{\tI,\tA'}
\end{equation}
\end{definition}
\begin{figure}[hbt]
 \setlength{\unitlength}{800sp}
    \begin{picture}(8745,3219)(931,-3565)
      {\thicklines \put(5401,-1261){\oval(1756,1756)}}
      {\put(1801,-1261){\line(1, 0){2700}}}
      {\put(6301,-1261){\vector(1, 0){3300}}}
      {\put(1801,-3361){\vector(1, 0){7800}}}
      \put(2026,-2611){\makebox(0,0)[b]{$\Phi$}}
      \put(5401,-1486){\makebox(0,0)[b]{$\tA$}}
      \put(9676,-2811){\makebox(0,0)[b]{$\Phi_{\tA,\tI}$}}
    \end{picture}
  \setlength{\unitlength}{800sp}
    \begin{picture}(8745,3219)(931,-3565)
      {\thicklines \put(5401,-3461){\oval(1756,1756)}}
      {\put(1801,-3361){\line(1, 0){2700}}}
      {\put(6301,-3361){\vector(1, 0){3300}}}
      {\put(1801,-1261){\vector(1, 0){7800}}}
      \put(2026,-2611){\makebox(0,0)[b]{$\Phi$}}
      \put(5401,-3686){\makebox(0,0)[b]{$\tA'$}}
      \put(9676,-2811){\makebox(0,0)[b]{$\Phi_{\tI,\tA'}\equiv\Phi_{\tA,\tI}$}}
    \end{picture}
\caption{\small Illustration of the concept of {\em twin} involution.}
\end{figure}
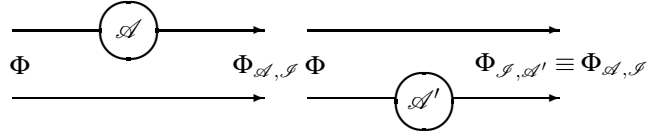
Notice that, by definition, independently on the faithful state $\Phi$ we always have trivially
\begin{equation}
\tI'=\tI.
\end{equation}
We now derive the Bloch matrix representation of the twin involution. The bipartite state in the
Bloch form is represented by the matrix
\begin{equation}
F_{ij}\doteq n_i\odot n_j(\Phi).\label{FBloch}
\end{equation}
The matrix $F$ is real and invertible, as a consequence of faithfulness of state $\Phi$ (by
definition the correspondence $\tilde\Phi_{\tA,\tI}\leftrightarrow \tA$ is one-to-one).
Indeed, a transformation $\tA$ on the first system is described by the matrix multiplication
\begin{equation}
n_i\odot n_j(\tilde\Phi_{\tA,\tI})=\sum_k A_{ik}F_{kj}=(AF)_{ij},
\end{equation}
where $A\doteq M(\tA)$. On the other hand, a transformation $\tA$ on the second system is
represented as
\begin{equation}
n_i\odot n_j(\tilde\Phi_{\tI,\tA})=\sum_k A_{jk}F_{ik}=(F\transp{A})_{ij}.
\end{equation}
One can also check the composition rules
\begin{eqnarray}
n_i\odot n_j(\tilde\Phi_{\tB\circ\tA,\tI})&=&(BAF)_{ij},\label{BAF}\\
n_i\odot n_j(\tilde\Phi_{\tI},\tB\circ\tA)&=&(F\transp{A}\transp{B})_{ij}\equiv
(F\transp{(BA)})_{ij}.\label{FAB}
\end{eqnarray}
Also, if one considers another faithful state $\Psi$ which is obtained by applying an invertible
deterministic transformation $\map{M}$ to the first system in the joint state $\Phi$, namely
\begin{equation}
\Psi=\Phi_{\tM,\tI},
\end{equation}
then the matrix $F$ in Eqs. (\ref{BAF}) and (\ref{FAB}) is substituted by the matrix $MF$. 
The defining identity (\ref{twinid}) now corresponds to the matrix identity
\begin{equation}
AF=F\transp{(A')},
\end{equation}
namely the twin involution is given by
\begin{equation}
A'=\transp{F}\transp{A}\transp{F}{}^{-1}.
\end{equation}
If the faithful state $\Phi$ is also {\em symmetric}, the twin involution satisfies all four axioms
of {\em generalized adjoint}:  
\begin{definition}[Generalized adjoint]\label{d:genad}
1. $(A+B)'= A'+B'$, 2. $(A')'=A$, 3. $(AB)'= B' A'$, 4. $A' A=0\Longrightarrow A=0$.
\end{definition}
Indeed, a faithful symmetric state has a Bloch representation in terms of a symmetric matrix $F$ in
Eq. (\ref{FBloch}). Therefore, the first three axioms are obvious. We just need to check the last
axiom. For this purpose we need the following simple lemma 
\begin{lemma}\label{lAtA} The following implication holds
\begin{equation}
\transp{A}A=0\Longrightarrow A=0.
\end{equation}
\end{lemma}
\Proof Using the real polar decomposition $A=PR$, with $P\geq 0$ positive symmetric and
$R\transp{R}=\transp{R}R=I$ (rotation matrix), one has that $\transp{A}A=\transp{R}P^2R$ has all
positive eigenvalues, each one is the square of the corresponding eigenvalue of $P$, whence
$\transp{A}A=0$ if all eigenvalues of $P$ are zero, namely $P=0$, or, equivalently, $A=PR=0$, since
$R$ is invertible.\qed \medskip 
\par We can now check that axiom 4. for the real adjoint holds for symmetric $F$, namely Postulate
\ref{p:faith} implies the existence of a {\em transposition} (the real equivalent of the {\em
  adjoint}), which can be operationally defined via the twin involution on a faithful symmetric state.
\medskip
\begin{theorem}[Operational adjoint] The existence of a symmetric faithful bipartite states
  guarantees the existence of a transposition on the real algebra $\aA$ of transformations.
\end{theorem}
\Proof Suppose that there exists a symmetric faithful state $\Phi$. Its matrix $F$ is symmetric
invertible. Then also $F^{-1}$ is symmetric. By real polar decomposition of $A$, we write
\begin{equation}
A' A=\transp{F}\transp{R}PF^{-1/2}F^{-1/2}PR,
\end{equation}
and invertibility of $F$ implies that $A' A=0$ is equivalent to
\begin{equation}
\transp{R}PF^{-1/2}F^{-1/2}PR=0,
\end{equation}
and using Lemma \ref{lAtA} one has
\begin{equation}
F^{-1/2}PR=0,
\end{equation}
namely $A=PR=0$. This proves identity 4. of Definition \ref{d:genad}, completing the list of
requirements that the twin involution must satisfy in order to be a generalized adjoint.\qed
\begin{lemma} For a faithful symmetric state $\Phi$ the following identities hold
\begin{equation}
\tilde\Phi_{\tA,\tB}=\tilde\Phi_{\tI,\tB\circ\tA'}=\tilde\Phi_{\tA\circ\tB',\tI}.
\end{equation}
\end{lemma}
\Proof
\begin{equation}
\tilde\Phi_{\tA,\tB}=(\tilde\Phi_{\tA,\tI})_{\tI,\tB}=(\tilde\Phi_{\tI,\tA'})_{\tI,\tB}=
\tilde\Phi_{\tI,\tB\circ\tA'}=\tilde\Phi_{\tA\circ\tB',\tI}.
\end{equation}
\qed
\begin{definition}[Real positive form]\label{defposform}
A linear form $\varphi$ over the algebra of transformations
  $\aA$ is called real positive (with respect to the real adjoint $\tA\to\tA'$) if 
$\forall\tA\in\aA$ it  satisfies the following identities
\begin{itemize}
\item[a)] $\varphi(\tA')=\varphi(\tA)$,
\item[b)]  $\varphi(\tA'\circ\tA)\ge 0$.
\end{itemize}
\end{definition}
\begin{theorem} The local state $\Phi|_1=\Phi|_2$ of  a symmetric faithful state $\Phi$ is a real
  positive form over $\aA$.  
\end{theorem}
\Proof From identity (\ref{faith4}) we have that $\Phi|_1=\Phi|_2$. Condition a) also follows from
the same identity. On the other hand, the condition b) holds also for generalized transformations,
since a generalized transformation is always a multiple of a physical one by a real scalar. \qed
\subsection{Mackey-Kakutani (MK) construction of real Hilbert space structure}
In the following we will show how the existence of a generalized adjoint over transformations allows
us to derive a structure of real Hilbert space over generalized weights. For this purpose we need the following
two theorems by Mackey and Kakutani\cite{Kakutani-Mackey}. 
\begin{theorem}[Mackey-Kakutani I]\label{t:km1} [Ref. \cite{Kakutani-Mackey}]. Let ${\mathfrak B}$
  be a real Banach space, and ${\mathfrak R}$ the ring of continuous linear transformations of
  ${\mathfrak B}$ into itself. Then ${\mathfrak B}$ is isomorphic to a (generally non separable) real Hilbert
  space $\sH$ if and only if there is an operation $\map{T}\to\map{T}'$ from ${\mathfrak R}$ to
  ${\mathfrak R}$ which has the properties of definition (\ref{d:genad}).
\end{theorem}
\begin{theorem}[Mackey-Kakutani II]\label{t:km2} [Ref. \cite{Kakutani-Mackey}].
  The isomorphism in Theorem \ref{t:km1} may be set up in such a manner that the correspondence
  $\map{T}\to\map{T}'$ goes over into the correspondence between its operator and its adjoint. In
  other words, ${\mathfrak B}$ may be provided with a positive definite symmetric bilinear inner
  product $(x,y)$ such that the new norm $\nn{x}$ in ${\mathfrak B}$ defined by the equation
  $\nn{x}=\sqrt{(x,x)}$ is equivalent to the given norm $\n{x}$ and such that for all $x$ and $y$ in
  ${\mathfrak B}$ $(T(x),y)=(x,T'(y))$.
\end{theorem}
\bigskip
\par Theorems \ref{t:km1} and \ref{t:km2} entail the following Hilbert space formulation:
\begin{remark}[Hilbert space structure for the Banach space of  generalized weights] Take for
  ${\mathfrak B}$ the Banach space of  generalized weights $\Wset$ and for ${\mathfrak R}$ the ring of linear
  transformations of $\Wset$ according to Eq. (\ref{wtransf}). Then Theorems \ref{t:km1} and
  \ref{t:km2} assert that the space of  generalized weights $\Wset$ is isomorphic to a real Hilbert space $\sH$,
  and that it is possible to choose the scalar product in such a way that the twin transform
  corresponds to the real-adjoint---\ie the transposition---and the norm is equivalent to the one
  induced by the scalar product. The Riesz theorem implies that the affine space of generalized propensities
  (linear real forms over states or, equivalently, over  generalized weights) is itself a real Hilbert space
  isomorphic to $\sH$.
\end{remark}
\subsection{Gelfand-Naimark-Segal (GNS) construction of real Hilbert space structure} With the
introduction of a generalized adjoint given in Definition in \ref {d:genad} corresponding to the
operational concept of twin involution, the real algebra $\aA$ of generalized transformations becomes a real
${}^*$-algebra. Then each real positive form $\varphi$ over the ${}^*$-algebra $\aA$---\eg the local
state $\varphi\doteq\Phi|_1$ of a faithful symmetric state $\Phi$---defines a Hilbert space
$\sH_\varphi$ and a representation $\pi_\varphi$ of $\aA$ by linear operators acting on
$\sH_\varphi$. Indeed, $\aA$ is a linear space over $\Reals$ and $\varphi$ defines a symmetric
(positive semi-definite) scalar product on $\aA$ as follows
\begin{equation}
{}_\varphi\!\<\tA|\tB\>_\varphi\doteq\varphi(\tA'\circ\tB)\equiv\Phi(\tA',\tB'),
\qquad \tA,\tB\in\aA,\label{scalprod}
\end{equation}
where we remind the use of notation defined in Eq. (\ref{notlocal}).
Indeed, condition a) of Definition \ref{defposform} implies the symmetry ${}_\varphi\!\<\tB|\tA\>_\varphi={}_\varphi\!\<\tA|\tB\>_\varphi$,
whereas condition b) implies the positivity ${}_\varphi\!\<\tA|\tA\>_\varphi\ge 0$.  Also, it is easy to check that
\begin{equation}
{}_\varphi\!\<\tC'\circ\tA|\tB\>_\varphi={}_\varphi\!\<\tA|\tC\circ\tB\>_\varphi,
\end{equation}
as it can be derived from the definition (\ref{scalprod}) as follows
\begin{equation}
{}_\varphi\!\<\tC'\circ\tA|\tB\>_\varphi=\Phi(\tA'\circ\tC,\tB')=
\tilde\Phi_{\tC,\tI}(\tA',\tB')=\tilde\Phi_{\tI,\tC'}(\tA',\tB')=
\Phi(\tA',\tB'\circ\tC')={}_\varphi\!\<\tA|\tC\circ\tB\>_\varphi
\end{equation} 
Symmetry and positivity imply the bounding
\begin{equation}
{}_\varphi\!\<\tA|\tB\>_\varphi\le\sqrt{{}_\varphi\!\<\tA|\tA\>_\varphi{}_\varphi\!\<\tB|\tB\>_\varphi}.\label{boundscal}
\end{equation}
Using the bounding (\ref{boundscal}) for the scalar product ${}_\varphi\!\<\tA'\circ\tA\circ\tX|\tX\>_\varphi$
we can easily see that the set $\aI\subseteq\aA$ consisting of all elements $\tX\in\aA$ with
$\varphi(\tX'\circ\tX)=0$ is a left ideal, \ie a linear subspace of $\aA$ which is stable under
multiplication by any element of $\aA$ on the left (\ie $\tX\in\aI$, $\tA\in\aA$ implies
$\tA\circ\tX\in\aI$).
The set of equivalence classes $\aA/\aI$ thus becomes a real pre-Hilbert
space equipped with a symmetric scalar product, an element of the space being an equivalence
class. Notice that the scalar product does not depend on the algebraic representatives chosen for
classes, namely
\begin{equation}
{}_\varphi\!\<\{\tA\}|\{\tB\}\>_\varphi={}_\varphi\!\<\tA|\tB\>_\varphi,\quad\forall\tA\in\{\tA\},\;\forall\tB\in\{\tB\},
\end{equation}
$\{\tA\}$ denoting the equivalence class containing $\tA$. For the equivalence classes we can define
the norm
\begin{equation}
\n{\tX}_\varphi^2\doteq{}_\varphi\!\<\tX|\tX\>_\varphi,\qquad \tX\in\aA/\aI.\label{normGNS}
\end{equation}
We keep the subindex $\varphi$ for the norm in order to distinguish it from the previously defined
norm (\ref{norm}). The Hilbert space is then obtained by completion of $\aA/\aI$ in the norm
topology (the Hilbert space closure is not operationally relevant: see Remark \ref{r:closure}).
 The product in $\aA$ defines the action of $\aA$ on the vectors in $\aA/\aI$, by
associating to each element $\tA\in\aA$ the linear operator $\pi_\varphi(\tA)$ defined on the dense
domain $\aA/\aI\subseteq\sH_\varphi$ as follows
\begin{equation}
\pi_\varphi(\tA)|\tX\>_\varphi\doteq|\{\tA\circ\tB\}\>_\varphi,\quad \tX=\{\tB\}.
\end{equation}
The norm (\ref{normGNS}) can be extended to a seminorm on the whole $\aA$ as follows 
\begin{equation}
\n{\tA}_\varphi\doteq\n{\{\tA\}}_\varphi=\sqrt{{}_\varphi\!\<\{\tA\}|\{\tA\}\>_\varphi}.\label{normGNSext}
\end{equation}
On the other hand, on $\aA/\aI$ one can easily verify that $\n{\cdot}_\varphi$ indeed satisfies all
axioms of norm, since clearly $\n{\tA}_\varphi=0$ implies that $\tA\in\aI$, corresponding to the
null vector, and
\begin{equation}
\begin{split}
&\n{\{\lambda\tA\}}_\varphi=\n{\lambda\{\tA\}}_\varphi
=\lambda\n{\{\tA\}}_\varphi,\\
&\n{\{\tA+\tB\}}_\varphi=\n{\{\tA\}+\{\tB\}}_\varphi\le\n{\{\tA\}}_\varphi+\n{\{\tB\}}_\varphi.
\end{split}
\end{equation}
If $\aA$ were a Banach ${}^*$-algebra the domain of definition of $\pi_\varphi(\tA)$ could be easily extended
to the whole $\sH_\varphi$ by continuity, since to a Cauchy sequence $\tX_n\in\aA/\aI$ there
correspond Cauchy sequences $\tA\tB_n$, $\tB_n\in\tX_n$ as a consequence of the norm  bounding 
\begin{equation}
\n{\pi_\varphi(\tA)\tX_n-\pi_\varphi(\tA)\tX_m}_\varphi=
\n{\{\tA(\tB_n-\tB_m)\}}_\varphi=\n{\tA(\tB_n-\tB_m)}_\varphi\le
\n{\tA}_\varphi\n{\tB_n-\tB_m}_\varphi.
\end{equation}
However, the last step is not necessarily true, since conditions 
$\n{\tB\circ\tA}_\varphi\le\n{\tB}_\varphi\n{\tA}_\varphi$, and $\n{\tA'}_\varphi=\n{\tA}_\varphi$
do not necessarily hold, whence the possibility of representing generalized transformations as operators
over $\sH_\varphi$ remains an open problem for the infinite dimensional case.  Also, the use of the
seminorm (\ref{seminorm}) closure is not of much help, since one can just prove that
\begin{equation}
\n{\tA}_\varphi\le\n{\tA'},\qquad\n{\tA}_\varphi^2\le\n{\tA'}\n{\tA},\label{bnorms}
\end{equation}
but we cannot prove a bounding $\n{\tB}\le\n{\tX}_\varphi$, $\tB\in\tX$. 
The first bound in Eq. (\ref{bnorms}) can be derived as follows
\begin{equation}
\n{\tA}_\varphi=\Phi(\tA',\tA')=\Phi_{\tI,\tA'}(\tA',\tI)\Phi(\tI,\tA')=\Phi_{\tI,\tA'}|_1(\tA')\Phi|_2(\tA')
\le\n{\tA'}^2,
\end{equation}
where $\Phi$ is any faithful state corresponding to $\varphi$. The second bound in
Eq. (\ref{bnorms}) is implied by the inequality
\begin{equation}
\n{\tA}_\varphi^2=\varphi(\tA'\circ\tA)\le\n{\tA'\tA}\le\n{\tA'}\n{\tA}.
\end{equation}
Also we do not have that $\n{\tA'}=\n{\tA}$, not even $\n{\tA'}\le\n{\tA}$.
\par In terms of the faithful state $\Phi$ and of its Bloch representation the scalar
  product (\ref{scalprod})  rewrites as
\begin{equation}\label{scalprod2}
{}_\varphi\!\<\tA|\tB\>_\varphi=\Phi(\tA',\tB')=(A'F\transp{B'})_{00}=(\transp{F}\transp{A}\transp{F}{}^{-1}BF)_{00}.
\end{equation}
\begin{remark}[Pairing between states and propensities]
From the definition (\ref{scalprod}) of the scalar product we have
\begin{equation}
{}_\varphi\!\<\tA'|\tB\>_\varphi=\varphi_\tB(\tA)\varphi(\tB)=\Phi_{\tI,\tB'}|_1(\tA),
\end{equation}
and if we assume that the state $\Phi$ is preparationally faithful, then for every state $\omega$
there exists a transformation $\tT_\omega$ such that $\omega=\Phi_{\tT_\omega,\tI}|_2=
\Phi_{\tI,\tT_\omega}|_1=\varphi_{\tT'_\omega}$ with $\varphi(\tT_\omega)\neq0$. Then one has
\begin{equation}\label{GNSw}
\omega(\tA)={}_\varphi\!\<\tA'|\tilde\tT_\omega\>_\varphi={}_\varphi\!\<\cA'|\tilde\tT_\omega\>_\varphi,\qquad
\tilde\tT_\omega=\frac{\tT'_\omega}{\varphi(\tT_\omega)}, 
\end{equation}
and we recover the pairing between states and propensities in terms of the scalar product. 
\end{remark}
Notice that state $\varphi$ is cyclic.
Eq. (\ref{GNSw}) along with the bounds in Eq. (\ref{bnorms}) imply the following theorem
\begin{theorem}\label{t:classes} Two (bounded) generalized transformations belong to the same
  equivalence class in $\aA/\aI$ if and only if they are informationally equivalent, namely
  $\tA\in\{\tB\}\Leftrightarrow\tA\in\cB$.
\end{theorem}
\Proof If $\tA$ is informational equivalent to $\tB$, then $\omega(\tA-\tB)=0$
$\forall\omega\in\Stset$, which implies that $\n{\tA-\tB}=0$, whence, according to the second bound
in Eq. (\ref{bnorms}), $\n{\tA-\tB}_\varphi=0$ if both $\tA$ and $\tB$ are bounded (for any
generalized transformation with bounded norm $\n{\tA}$ one has $\n{\tA'}<\infty$, since
one can write $\tA=\lambda\tT$, with $\tT$ a true transformation and $|\lambda|<\infty$, and $\tT'$
bounded, being $\tT'$ a true transformation by definition of the real adjoint). This means that
$\tA=\tB+\tX$, with $\tX\in\aI$, namely $\tA\in\{\tB\}$. Reversely, if $\tA\in\{\tB\}$, then one has
$\tA=\tB+\tX$, with ${}_\varphi\!\<\tX|\tX\>_\varphi=0$. Using Eq. (\ref{GNSw}) we have the bounding
\begin{equation}\label{smartbound}
\omega(\tX)={}_\varphi\!\<\tX'|\tilde\tT_\omega\>_\varphi\le\n{\tX}_\varphi\n{\tilde\tT_\omega}_\varphi,
\end{equation}
whence if $\n{\tX}_\varphi=0$, then $\omega(\tA-\tB)=\omega(\tX)=0$ for all states $\omega$, namely
$\tA$ is informationally equivalent to $\tB$.  \qed Therefore, the vectors of the Hilbert space
$\sH_\varphi$ are in one-to-one correspondence with generalized propensities.  From the bounding
(\ref{smartbound}) we can also see that if the state $\varphi$ satisfies
$\n{\tilde\tT_\omega}_\varphi\leq C_\varphi$ for some constant $C_\varphi\geq 0$ depending only on
  $\varphi$, then one can also reversely bound the two inequivalent norms $\n{\cdot}$ and
  $\n{\cdot}_\varphi$ as follows
\begin{equation}
\n{\tA}\le C_\varphi\n{\tA}_\varphi.
\end{equation}
In such case one the domain of definition of $\pi_\varphi(\tA)$ can be extended to the whole Hilbert
space $\sH_\varphi$.

\section{Dimensionality theorems}\label{s:dimensionality}
We will now consider the consequences of Postulates \ref{p:locobs} and \ref{p:Bell}. We will see
that they entail dimensionality theorems that agree with the tensor product rule for Hilbert spaces
for composition of independent systems in Quantum Mechanics. Moreover, Postulate \ref{p:Bell}, in
particular, shows that the real Hilbert space $\sH_\varphi$ is isomorphic to the real Hilbert space
of Hermitian complex matrices representing selfadjoint operators over a complex Hilbert space $\sH$ of
dimensions equal to $\idim{\Stset}$, finally leading to the Hilbert space formulation of Quantum Mechanics.

\par The {\em local observability principle} \ref{p:locobs} is operationally crucial, since it
reduces enormously the complexity of informationally complete observations on composite systems, by
guaranteeing that only local (although jointly executed!) experiments are sufficient for retrieving
a complete information, also any correlations between the component systems.  This principle
directly implies the following upper bound for the affine dimension of a composed system
\begin{equation}\label{admbound1}
\adm(\Stset_{12})\le\adm(\Stset_1)\adm(\Stset_2)+\adm(\Stset_1)+\adm(\Stset_2).
\end{equation}
In fact, if the number of outcomes of a minimal informationally complete observable on $\Stset$ is
$N$, the affine dimension is given by $\adm(\Stset)=N-1$ (since the number of outcomes must equal
the dimension of the affine space embedding the convex set of states $\Stset$ plus another dimension
for the normalization functional $n_0$). Now, consider a global informationally
complete measurement made of two local minimal informationally complete observables measured
jointly. It has number of outcomes $[\adm(\Stset_1)+1][\adm(\Stset_2)+1]$. However, we are not
guaranteed that the joint observable is itself minimal, whence the bound (\ref{admbound1}) follows.
\par We now translate the concept of dynamically faithful state in the Bloch
representation. If the state $\Phi$ is (dynamically) faithful, then the output state
$\Phi_{\tA,\tI}$ (conditioned that the transformation $\tA$ occurred locally on the first system) is
in one-to-one correspondence with the transformation $\tA$.  Therefore, one can completely determine
the transformation by determining the output state. We need to determine the matrix $\vec M(\tA)$
plus the vectors $\vec k(\tA)$ and $\vec m(\cA)$, plus the parameter $q(\cA)$, namely
$\adm(\Stset)^2+2\adm(\Stset)+1$ parameters. However, one parameter, say $q(\cA)$ is determined by
the overall probability of occurrence of $\tA$ on the state $\Phi$, from which the conditioned state
is independent. Therefore, in order to have a joint faithful state we need to have at least
$\adm(\Stset)[\adm(\Stset)+2]$ independent parameters for the joint state, namely we have the lower
bound for the affine dimension of the joint system
\begin{equation}\label{admbound2}
\adm(\Stset^{\times 2})\ge\adm(\Stset)[\adm(\Stset)+2].
\end{equation}
If we put the two bounds (\ref{admbound1}) and (\ref{admbound2}) together, for a bipartite system
made of two identical systems  we obtain
\begin{equation}
\adm(\Stset^{\times 2})=\adm(\Stset)[\adm(\Stset)+2],
\end{equation}
which agrees with the dimensionality of composite systems in Quantum Mechanics coming from the
tensor product. The Bloch representation can be obtained experimentally by performing a joint
informationally complete measurement on both systems at the output, and then:
\begin{enumerate}
\item determining the probability of occurrence of the transformation $\tA$ on the state $\Phi$,
  which is given by
\begin{equation}
\Phi(\tA,\tI)=\Phi(\cX_0\circ\tA,\cX_0)=(\vec m(\cA)\cdot\vec n\odot n_0)(\Phi)+q(\cA);
\end{equation}
\item determining the following probabilities
\begin{equation}
\begin{split}
\Phi(\cX_j\circ\tA,\cX_k)=&\frac{[(\vec M(\tA)\vec n)_j\odot\vec n_k](\Phi)+\vec k_j(\tA)
(n_0\odot\vec n_k)(\Phi)}{\Phi(\tA,\tI)},\qquad 
\begin{matrix}
j=1,\ldots\adm(\Stset),\\k=0,1,\ldots\adm(\Stset),
\end{matrix}\\
\Phi(\cX_0\circ\tA,\cX_j)=&(\vec m(\cA)\cdot\vec n\odot\vec n_j)(\Phi)+q(\cA),
\qquad j=1,\ldots\adm(\Stset);
\end{split}
\end{equation}
\item invert the above equations in terms of $\vec M(\tA)$, $\vec k(\tA)$, $\vec m(\cA)$, and $q(\cA)$.
\end{enumerate}
Assuming now Postulate \ref{p:Bell} gives a bound for the informational dimension of the
informational dimension of convex sets of states. In fact, if for any bipartite system made of two
identical components and for some preparations of one component there exists a discriminating
observable that is informationally complete for the other component, this means that
$\adm(\Stset)\ge\idim{\Stset^{\times 2}}-1$, with the equal sign if the informationally complete
observable is also minimal, namely
\begin{equation}\label{infcomfromdiscr}
\adm(\Stset)=\idim{\Stset^{\times 2}}-1.
\end{equation}
By comparing this with the affine dimension of the bipartite system, we get
\begin{equation}
\adm(\Stset^{\times 2})=\adm(\Stset)[\adm(\Stset)+2]=[\idim{\Stset^{\times 2}}-1][\idim{\Stset^{\times 2}}+1]=\idim{\Stset^{\times 2}}^2-1,
\end{equation}
which, generalizing to any convex set gives the identification
\begin{equation}\label{Hd}
\adm(\Stset)=\idim{\Stset}^2-1,
\end{equation}
corresponding to the dimension of the quantum convex sets $\Stset$ originated from Hilbert spaces.
Moreover, upon substituting Eq. (\ref{infcomfromdiscr}) into Eq. (\ref{Hd}) one obtain
\begin{equation}\label{Hd2}
\idim{\Stset^{\times 2}}=\idim{\Stset}^2,
\end{equation}
which is the {\em tensor product rule} for informational dimensionalities.
\par According to Theorem \ref{t:classes} we have the identity
\begin{equation}\label{dimWset}
\dim(\sH_\varphi)=\adm(\Stset)+1,
\end{equation}
since $\sH_\varphi$ is identified with the vector space of the generalized propensities, namely the
space of the linear functionals over states which has one more dimension than the convex set of
states corresponding to normalization. From Eqs. (\ref{Hd}) and (\ref{dimWset}) we now have
\begin{equation}
\dim(\sH_\varphi)=\idim{\Stset}^2.
\end{equation}
Then, for finite dimensions the real Hilbert space $\sH_\varphi$ is isomorphic to the real Hilbert
space of Hermitian complex matrices representing selfadjoint operators over a complex Hilbert space
$\sH$ of dimensions $\dim(\sH)=\idim{\Stset}$, with scalar product corresponding to the trace pairing
used in the Born rule, and with the convex cones of propensities and states corresponding to the
convex cone of positive matrices. This is the Hilbert space formulation of Quantum Mechanics. In
infinite dimensions the selfadjoint operators are generally unbounded, since norm $\n{\cdot}$ is not
necessarily bounded, and boundedness of probabilities is provided by the faithful state $\Phi$.
\bigskip
\par In deriving Eq. (\ref{Hd}) I have implicitly assumed that the relation between the affine
dimension and the informational dimension which holds for bipartite systems must hold for any
system. Indeed, one can prove independently that 
\begin{equation}
\idim{\Stset^{\times 2}}\geq\idim{\Stset}^2,
\end{equation}
since locally perfectly discriminable states are also jointly discriminable, and the existence of a
preparationally faithful state guarantees the existence of $\idim{\Stset}^2$ jointly discriminable
states, the bound in place of the identity coming from the fact that we are not guaranteed that the
set of jointly discriminable states made of local ones is maximal. At the present stage of this
research in progress it is still not clear if the mentioned implicit assumption is avoidable, and,
if not, how relevant it is. One may need to add another postulate requiring a kind of universality
of informational laws--- such as $\adm(\Stset)=\idim{\Stset}^2-1$---independently on the physical
system, \ie on the convex set of states $\Stset$. It is also possible that in this way Postulate
\ref{p:Bell} can be avoided. These issues will be analyzed in detail in a forthcoming publication.

\section*{Acknowledgments}
This research has been completely supported from the Italian Minister of University and Research
(MIUR) under programs Prin 2003, Prin 2005 and Firb (bando 2001). The work has been feasible
part-time during my summer and Christmas visits in 2004 and 2005 at Northwestern University, thanks
to the kind hospitality of Horace Yuen. I wish to thank Lucien Hardy, Chris Fuchs, Reinhard Werner,
and Alexander Holevo for interesting and stimulating discussions, Guido Bacciagaluppi and Jos Uffink
for a useful analysis of a preliminary version of this work, Gregg Jaeger and Karl Svozil for
valuable suggestions on the linguistic side. A special thank to Giulio Chiribella, Paolo Perinotti,
and Massimiliano Sacchi, for their invaluable critical analysis of the manuscript.  Finally, a
particular thank to Maria Luisa Dalla Chiara for her encouraging enthusiastic support.

\end{document}